\documentclass{ws-procs975x65}
\newcommand{\be}{\begin{equation}}
\newcommand{\ee}{\end{equation}}
\newcommand{\bea}{\begin{eqnarray}}
\newcommand{\eea}{\end{eqnarray}}
\begin{document}


\title{A semiclassical approach to $\eta/s$ bound through holography}

\author{Alessandro Pesci\footnote{pesci@bo.infn.it}}
\address{INFN, Sezione di Bologna, Via Irnerio 46, I-40126 Bologna, Italy}


\begin{abstract}
We consider the holographic principle, in its lightsheet formulation, in the 
semiclassical context of statistical-mechanical systems in classical Einstein 
spacetimes. A local condition, in terms of entropy and energy local densities  
of the material medium under consideration, is discussed, which turns out to 
be necessary and sufficient for the validity of the closely-related 
generalized covariant entropy bound. This condition is apparently a general 
consequence or expression of flat-spacetime quantum mechanics alone, without 
any reference to gravity. Using it, a lower bound $\eta/s \geq 1/4\pi$ can be 
derived, with the limit attained (in certain circumstances) by systems 
hydrodynamically dominated by  radiation quanta.\\
\end{abstract}

\bodymatter

For thermal theories with holographic gravitational
dual,
certain quantities, albeit unrelated to gravity, are however easily evaluated
through string calculations in the equivalent dual gravitational theory.
As far as the string calculations are performed in the semiclassical
limit, they could be expected to be readable also in a conventional 
quantum approach,
without reference to strings, for thermal systems at circumstances 
(to be defined) embodying the onset of the holographic duality above.
The Kovtun-Son-Starinets (KSS) bound \cite{KSS},
could be precisely an example of this. 
It says that for a large class of thermal field theories, 
even widely different from each other
but always with gravity duals,
$\eta/s = 1/4\pi$ (in Planck units, the units we use in this note), 
being also conjectured this to be in 
general a lower bound, at least for relativistic systems.

Now, holography is summarized, in the semiclassical
context of material media living in the continuous Einstein spacetime,
by the so-called generalized covariant entropy 
bound \cite{FMW} (genB).
This bound turns out to be universally true {\it iff} a lower limit is
definitely put to the scale $l$ 
of the statistical-mechanical description
for the assigned local conditions,
with this limit being apparently anyway 
required by the intrinsic space (and time \cite{Pesci3})
quantum uncertainty of the constituent
particles: \cite{Pesci1,Pesci2} 

\begin{eqnarray}\label{lstar}
l \geq
\lambda \geq
\frac{1}{\pi} \frac{s}{\rho+p} \equiv
l^*,
\end{eqnarray}
where $s$, $\rho$, $p$ are local entropy and energy density and pressure
respectively and $\lambda$ is the typical wavelength of
the constituent particles. The genB can be saturated 
(for peculiar geometric configurations) by systems
which have $\lambda = l^*$, i.e. the most entropic ones
(typically, ultrarelativistic gases).\cite{Pesci2}

Exploiting the notion of $l^*$ we have just introduced
we can calculate $\eta/s$ (along the lines of Ref.~\refcite{Pesci4}). 
For nonrelativistic systems, 
for example a gas at ordinary conditions,
we have 
$\eta = \frac{1}{3} L \rho a$ with $L$ the correlation length
and $a$ the thermal velocity of the constituent particles,
so that

\begin{eqnarray}\label{nr1}
\frac{\eta}{s} =
\frac{1}{3} \frac{L \rho a}{s} \simeq
\frac{1}{3} L \frac{\rho + p}{s} a =
\frac{1}{3 \pi} \frac{L}{l^*} a,
\end{eqnarray}
where use is made of the expression (\ref{lstar}) for $l^*$.

Now, if $\rho$, $s$, $p$ are allowed to increase
while $\frac{\rho+p}{s}$ and $\lambda$ are held fixed (for a gas this means
to increase number density with intensive parameters fixed),
$L$ decreases, but without going below its quantum mechanical
limit $\lambda$. If we assume that this limiting condition can actually
be reached we have

\begin{eqnarray}\label{nr2}
\left( \frac{\eta}{s} \right)_{min} =
\frac{1}{3\pi} \frac{\lambda}{l^*} a.
\end{eqnarray} 
Here $a$
can be also very near to 0 so that
the KSS bound could be violated.
However, considering explicitly the case of a Boltzmann gas
we see this is not the case. 
We have in fact for it \cite{Pesci2} $\frac{\lambda}{l^*} \gg 1$
with

\begin{eqnarray}\label{nr3}
\frac{\lambda}{l^*} =
(2\pi)^{3/2} \sqrt{\frac{m}{T}} \ \frac{1}{\chi+\frac{f+2}{2}} \propto
\sqrt{\frac{m}{T}} \propto
\frac{1}{a},
\end{eqnarray}
so that in $(\eta/s)_{min}$ no dependence on $a$ is left.
In this expression $m$ is the mass of constituent particles,
$\chi$ is a number $\simeq 0$ when conditions are such that
$L \simeq \lambda$, $f$ is the number of degrees of freedom per particle,
and use is made of the relation between temperature and thermal velocity,
$T = \frac{1}{3} m a^2$.
Putting numbers in, if we assume $f =2$ we find 
$(\eta/s)_{min} \approx 0.7$.
Thus the KSS bound seems can be satisfied also for nonrelativistic
systems, no matter how small $a$ can be.

Let us consider ultrarelativistic systems
consisting of interacting radiation.
We can have in mind 
massive particles with statistical equilibrium determined
by collisions with radiation quanta
(the particles only act as mechanism to transfer momentum through
the radiation field)
or directly a gas of photons
and ultrarelativistic electrons and positrons,
or gluons interacting among themselves and 
with ultrarelativistic quarks.
Assuming $\eta \approx \frac{1}{3} \tau \rho_{\gamma}$ \cite{Misner},
where $\tau$ is the average time for a quantum to collide
and $\rho_{\gamma}$ is the energy density of radiation,
we get

\begin{eqnarray}\label{r1}
\frac{\eta}{s_{\gamma}} \approx
\frac{1}{3} \tau \frac{\rho_\gamma}{s_\gamma} =
\frac{1}{4} \tau \frac{\rho_\gamma + p_\gamma}{s_\gamma} =
\frac{1}{4\pi} \frac{\tau}{l^*_\gamma} =
\frac{1}{4 \pi} \frac{L}{l^*_\gamma},
\end{eqnarray}
with $L$ the average distance for a quantum to collide
and $s_{\gamma}$ the entropy density of radiation.
Now, for a gas of radiation quanta still $L$ cannot be smaller
than its quantum limit $\lambda_\gamma$.
If we assume that, for assigned $T$, $L$ can decrease
down to the limit $L \rightarrow \lambda_\gamma = L_{min}$
(which, actually, amounts to imply strong coupling,
since, at the limiting conditions 
of one potential collision in every $\lambda_\gamma$,
$L = \frac{1}{\sigma n} \approx \frac{\lambda_\gamma}{g^2}$,
where $n$ is number density ($= \lambda_\gamma^{-3}$ for radiation) and
for $\sigma$
the expression of Thompson cross section is used
in terms of the coupling constant $g$), we get

\begin{eqnarray}\label{r2}
\left( \frac{\eta}{s} \right)_{min} \approx
\frac{1}{4\pi} \frac{\lambda_\gamma}{l_\gamma^*} =
\frac{1}{4\pi},
\end{eqnarray}
where last equality comes from the fact that for radiation
$\lambda = l^*$, from what mentioned above.

Apparently, the KSS limit must be there,
and the conditions for it to be attained are
the fluid has $\lambda = l^*$
(an ``holographic'' fluid, i.e. able to attain the generalized
covariant entropy bound) and $L = \lambda$.  
We should thus define the ``holographic duality conditions'' for our
semiclassic fluids to be $\lambda = l^* = L$.

As far as the QCD matter produced
at RHIC
can be described in terms of a fluid of strongly coupled radiation quanta,
the very low values of $\eta/s$ found at RHIC \cite{RHIC}
could thus find an explanation in what we have seen.

In conclusion,
to the value $\eta/s = 1/4\pi$ it is possible to arrive
both through string theoretical calculations in the gravity dual
of holographic thermal theories and, we have seen,
through ordinary flat-spacetime quantum arguments for fluids
at ``holographic duality'' $\lambda = l^* = L$.

In Ref.~\refcite{Hod2} a different non-string-theoretical argument 
bringing to the KSS limit is presented,
still relying on $l^*$ concept.


{\em Acknowledgements:} The author wishes to thank Paolo Benincasa
for fruitful discussions
and Shahar Hod for helpful correspondence.

\vfill

\end{document}